\pdfoutput=1
\documentclass[preprint,showpacs,preprintnumbers,amsmath,amssymb,floatfix,endfloats*]{revtex4}

\usepackage{graphicx}
\usepackage{dcolumn}
\usepackage{bm}
\usepackage{amsfonts}

\DeclareMathAlphabet{\mathsfsl}{OT1}{cmr}{bx}{it}
\begin{document}
\title{The influence of periodic shear on structural relaxation and pore redistribution in binary glasses}
\author{Nikolai V. Priezjev$^{1,2}$ and Maxim A. Makeev$^{3}$}
\affiliation{$^{1}$Department of Mechanical and Materials
Engineering, Wright State University, Dayton, OH 45435}
\affiliation{$^{2}$National Research University Higher School of
Economics, Moscow 101000, Russia}
\affiliation{$^{3}$Department of Chemistry, University of
Missouri-Columbia, Columbia, MO 65211}
\date{\today}
\begin{abstract}

The evolution of porous structure, potential energy and local
density in binary glasses under oscillatory shear deformation is
investigated using molecular dynamics simulations.  The porous
glasses were initially prepared via a rapid thermal quench from the
liquid state across the glass transition and allowed to phase
separate and solidify at constant volume, thus producing an extended
porous network in an amorphous solid. We find that under periodic
shear, the potential energy decreases over consecutive cycles due to
gradual rearrangement of the glassy material, and the minimum of the
potential energy after thousands of shear cycles is lower at larger
strain amplitudes. Moreover, with increasing cycle number, the pore
size distributions become more skewed toward larger length scales
where a distinct peak is developed and the peak intensity is
enhanced at larger strain amplitudes. The numerical analysis of the
local density distribution functions demonstrates that cyclic
loading leads to formation of higher density solid domains and
homogenization of the glass phase with reduced density.

\end{abstract}

\pacs{34.20.Cf, 68.35.Ct, 81.05.Kf, 83.10.Rs}


\maketitle

\section{Introduction}

The development of efficient strategies to synthesize hierarchically
structured porous materials with superior physical and mechanical
properties is important for a variety of applications including
adsorption, catalysis, separation, and medical devices~\cite{Su17}.
Recent atomistic and continuum simulation studies have demonstrated
that plastic deformation of metallic glasses with a regular array of
pores is facilitated by nucleation of shear bands at the pore
surfaces and their subsequent propagation along the planes populated
by
pores~\cite{Bargmann14,EckertMet15,EckertAct16,Gouripriya17,Jiang17,Song17,Luo18}.
It was recently found that under steady shear, the random porous
structure in binary glasses becomes significantly transformed and
dominated by a few large pores, and the shear modulus is a strong
function of porosity~\cite{Priezjev17s,Sahimi94}.  It was later
shown that tensile loading at constant volume leads to pore
coalescence and breaking of the glass in the regions of largest
spatial extent of reduced density~\cite{Priezjev18t}, while a nearly
homogeneous amorphous material can be formed during extended strain
at constant pressure~\cite{Priezjev18tp}.   Upon compression, the
pore shape is initially deformed, and, at large strain, adjacent
pores coalesce with each other, thus forming homogenous solid
domains that provide an enhanced resistance to
deformation~\cite{Priezjev18c}. However, the behavior of porous
glasses during more complex, time-dependent deformation protocols
remains relatively unexplored.

\vskip 0.05in

In recent years, atomistic simulations of amorphous solids under
cyclic shear were especially useful in elucidating molecular
mechanisms of the yielding transition, strain localization and
mechanical
annealing~\cite{Priezjev13,Sastry13,Reichhardt13,Priezjev14,IdoNature15,Priezjev16,Kawasaki16,Priezjev16a,
Sastry17,Priezjev17,OHern17,Hecke17,Keblinsk17,Priezjev18,Priezjev18a,Sastry18sh,NVP18strload}.
Here we highlight a few key results.  Particularly, it was shown
that below yield, binary glasses prepared with relatively large
quenching rates can be mechanically annealed toward lower potential
energy states, and the energy levels that can be accessed after
hundreds of shear cycles become deeper at larger strain
amplitudes~\cite{Sastry13,Sastry17,Priezjev18,
Priezjev18a,Sastry18sh,NVP18strload}.  In turn, the yielding
transition typically occurs after a number of transient cycles,
resulting in a formation of a shear band across the system, and, in
addition, the critical strain amplitude decreases with increasing
temperature~\cite{Sastry17,Priezjev17,Priezjev18a,Sastry18sh}. It
was also found that above the yielding point, periodic shear causes
irreversible deformation, plastic flow, and enhanced particle
diffusion, while the system is relocated to higher potential energy
states~\cite{Priezjev13,Sastry13,Reichhardt13,Priezjev14,IdoNature15,Priezjev16,
Kawasaki16,Priezjev16a,Sastry17,Priezjev17,OHern17,Hecke17,Keblinsk17,Priezjev18a,Sastry18sh}.
However, the effect of periodic deformation on structural and
mechanical properties of phase-separated systems including porous
glasses and gels remains to be fully understood.

\vskip 0.05in

In this paper, we perform molecular dynamics simulations to study
porous glasses subjected to oscillatory shear deformation.   The
porous glasses are formed during a coarsening process at low
temperature after a rapid thermal quench at constant volume. It will
be shown that in the absence of deformation, the potential energy
gradually decreases over time, the glass phase becomes denser, and
the pore size distributions are slightly skewed toward larger length
scales. With increasing strain amplitude, the evolution of the
porous structure after thousands of shear cycles becomes more
apparent, resulting in the formation of a pronouncing peak in the
distribution of larger pore sizes.

\vskip 0.05in

The rest of the paper is organized as follows. The preparation
procedure, deformation protocol, and parameter values are presented
in the next section.  The analysis of the porous structure and pore
size distributions as well as potential energy series and shear
stress are described in section\,\ref{sec:Results}. The results are
briefly summarized in the last section.

\section{MD simulations}
\label{sec:MD_Model}


The molecular dynamics simulations described in this section were
performed to study the dynamic response of a model glass to periodic
shear. We consider a binary mixture model (80:20) with strongly
non-additive interactions between different atom types that was
originally introduced by Kob and Andersen~\cite{KobAnd95}. In this
model, the interaction between two atoms of types $\alpha,\beta=A,B$
is specified via the Lennard-Jones (LJ) potential as follows:
\begin{equation}
V_{\alpha\beta}(r)=4\,\varepsilon_{\alpha\beta}\,\Big[\Big(\frac{\sigma_{\alpha\beta}}{r}\Big)^{12}\!-
\Big(\frac{\sigma_{\alpha\beta}}{r}\Big)^{6}\,\Big],
\label{Eq:LJ_KA}
\end{equation}
where $\varepsilon_{AA}=1.0$, $\varepsilon_{AB}=1.5$,
$\varepsilon_{BB}=0.5$, $\sigma_{AB}=0.8$, and $\sigma_{BB}=0.88$,
and $m_{A}=m_{B}$~\cite{KobAnd95}.  The cutoff radius is set to
$r_{c,\,\alpha\beta}=2.5\,\sigma_{\alpha\beta}$ to increase
computational speed. The following convention is adopted for the
reduced units of length, mass, energy, and time:
$\sigma=\sigma_{AA}$, $m=m_{A}$, $\varepsilon=\varepsilon_{AA}$, and
$\tau=\sigma\sqrt{m/\varepsilon}$, respectively. The total number of
atoms is $N=300\,000$.   The equations of motion were solved
numerically using the velocity-Verlet scheme~\cite{Allen87} with the
integration time step $\triangle t_{MD}=0.005\,\tau$~\cite{Lammps}.

\vskip 0.05in


We next briefly describe the preparation protocol for the porous
glass at constant volume, which is identical to the simulation
procedure used in the recent MD
studies~\cite{Kob11,Kob14,Priezjev17s,Makeev18,Priezjev18t,Priezjev18c,Priezjev18tp}.
The binary mixture was first thoroughly equilibrated at constant
volume and at the temperature of $1.5\,\varepsilon/k_B$ that is
larger than the glass transition temperature
$T_g\approx0.435\,\varepsilon/k_B$. Here, $k_B$ denotes the
Boltzmann constant.  In turn, the porous glass was formed during the
coarsening process at constant volume and temperature
$T_{LJ}=0.05\,\varepsilon/k_B$ after an instantaneous thermal quench
from the liquid state.  As a result, after the time interval of
$10^{4}\,\tau$, the porous structure is developed inside the
amorphous
material~\cite{Kob11,Kob14,Priezjev17s,Makeev18,Priezjev18t,Priezjev18c,Priezjev18tp}.
In the present study, due to limited computational resources, we
only considered one realization of disorder for a sample with the
average glass density $\rho\sigma^{3}=0.5$.


\vskip 0.05in

Following the preparation protocol, the porous glass was subjected
to oscillatory shear deformation at constant volume. More
specifically, the time-periodic strain was applied using the
following relation:
\begin{equation}
\gamma(t)=\gamma_{0}\,\,\textrm{sin}(2\pi t / T),
\label{Eq:strain}
\end{equation}
where $\gamma_{0}$ is the strain amplitude and $T$ is the
oscillation period.  To avoid confusion, in what follows, the
temperature is denoted by $T_{LJ}$, whereas the oscillation period
is referred to as $T$. We considered the values of strain amplitude
from $\gamma_{0}=0$ to $0.012$ and the oscillation period
$T=500\,\tau$. The simulations were carried out during 2000 shear
cycles for each value of $\gamma_{0}$. Note that the value
$\gamma_{0}=0$ corresponds to a quiescent sample that was aged at
constant volume during the time interval of $2000\,T=10^6\tau$. The
temperature $T_{LJ}=0.05\,\varepsilon/k_B$ was regulated using the
Nos\'{e}-Hoover thermostat~\cite{Lammps}. In addition, the
Lees-Edwards periodic boundary conditions were applied in the plane
of shear, while the usual periodic boundary conditions were imposed
along the vorticity direction~\cite{Lammps}. We comment that
considerable computational resources (hundreds of processors) were
required to simulate the system of $300\,000$ particles for
$2\times10^8$ MD steps. A relatively large system size was chosen to
avoid finite size effects~\cite{Kob11,Kob14}.   During production
runs, the potential energy, shear stress, pressure and temperature
as well as atomic configurations after each cycle were saved for the
postprocessing analysis.

\section{Results}
\label{sec:Results}


We begin the discussion by briefly reviewing the process of
formation of the porous glass and its properties after the initial
stage of phase separation.   It was previously shown that after an
instantaneous isochoric quench to a low temperature well below
$T_g$, the glass-forming liquid undergoes simultaneous phase
separation and solidification of the material~\cite{Kob11,Kob14}.
Moreover, the phase separation kinetics at sufficiently low
temperatures crosses over after about $10^2\,\tau$ from a power-law
increase to a logarithmically slow domain growth~\cite{Kob11,Kob14}.
The resulting porous structure after the time interval of $10^4\tau$
at $T_{LJ}=0.05\,\varepsilon/k_B$ for the sample with the average
glass density $\rho\sigma^{3}=0.5$ is typically characterized by a
percolating solid carcase and a complex porous
network~\cite{Makeev18}. It should be emphasized that the conditions
of constant volume and low temperatures generally result in the
formation of the porous structures at negative
pressures~\cite{Kob11,Kob14,Makeev18}. Remarkably, it was found that
the ratio of the pressure and temperature scales as
$P/T_{LJ}\sim\rho^{2.5}$ in a wide parameter range~\cite{Makeev18}.
In particular, the average value of the pressure for the glass
density $\rho\sigma^{3}=0.5$ at $T_{LJ}=0.05\,\varepsilon/k_B$ is
$P\approx-0.15\,\varepsilon/\sigma^3$, and the system is effectively
under tension in a confined geometry~\cite{Makeev18}.

\vskip 0.05in

\begin{figure}[t]
\includegraphics[width=12.cm,angle=0]{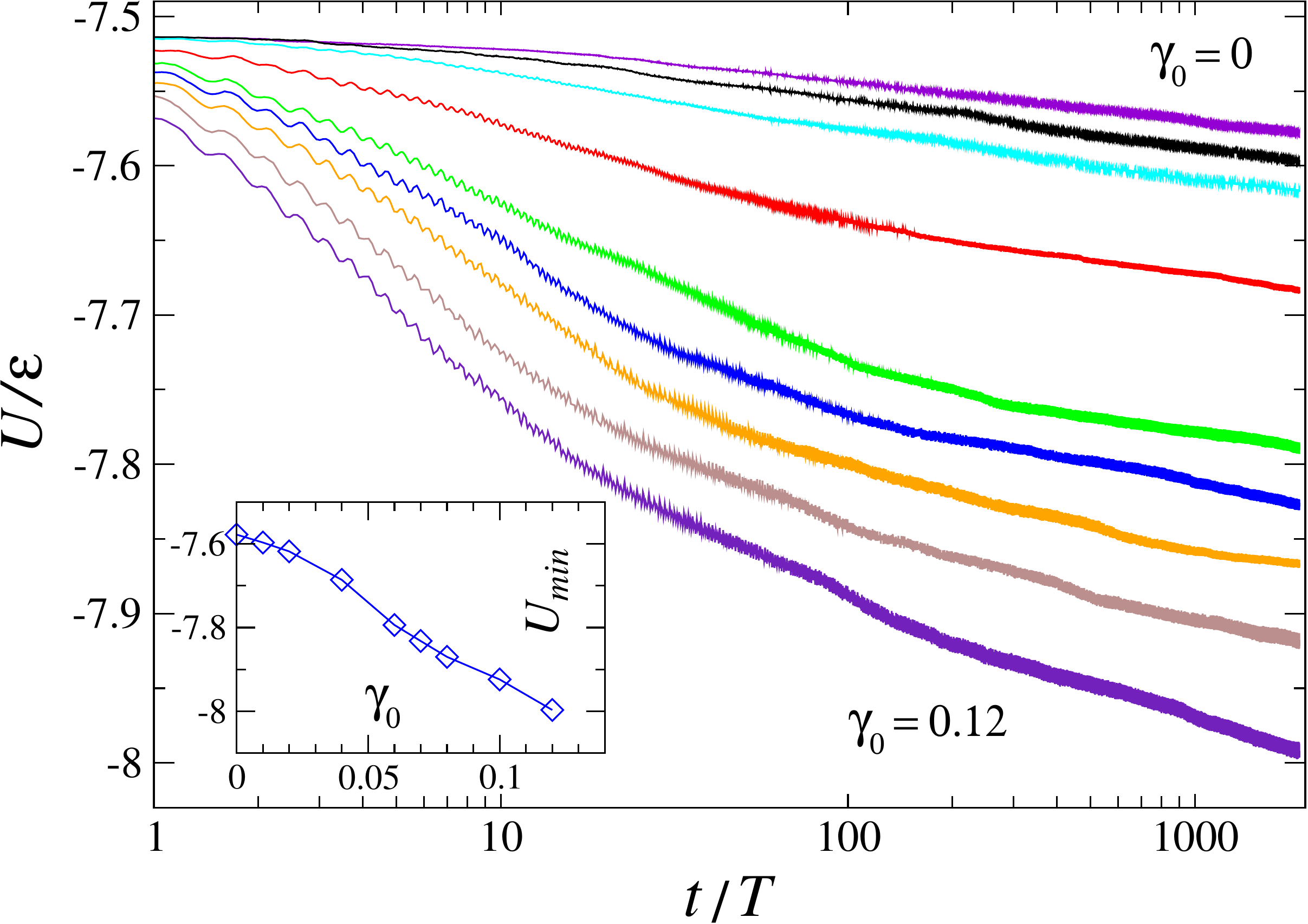}
\caption{(Color online) The potential energy per atom as a function
of time for strain amplitudes $\gamma_0=0$, $0.01$, $0.02$, $0.04$,
$0.06$, $0.07$, $0.08$, $0.10$, and $0.12$ (from top to bottom). The
oscillation period is $T=500\,\tau$ and the average glass density is
$\rho\sigma^{3}=0.5$. The inset shows the minimum of the potential
energy after 2000 cycles versus the strain amplitude.}
\label{fig:poten_cycle_number}
\end{figure}

The variation of the potential energy per atom as a function of
time, expressed in periods, is shown in
Fig.\,\ref{fig:poten_cycle_number} for different strain amplitudes.
When plotted on the log scale, it becomes evident that in each case
the potential energy gradually decreases over consecutive cycles,
and the steady states of deformation at the lowest energy levels are
yet to be reached.   Due to limited computational resources, we
examine the dynamic response to periodic shear in porous glasses
only during the first 2000 cycles.   Thus, the case $\gamma_{0}=0$
represents an aging process at constant volume in the undeformed
porous glass, which is essentially a continuation of the coarsening
dynamics extended to a time interval of $2000\,T=10^6\tau$. As will
be shown below, even in the absence of mechanical agitation, the
porous structure and the glass phase undergo a noticeable
transformation during $10^6\tau$. We comment that the relative
decrease in the potential energy at constant volume in the quiescent
sample (shown in Fig.\,\ref{fig:poten_cycle_number}) is
significantly smaller than the energy decrease during aging at
constant pressure, which occurs due to volume decrease and
densification of the glass phase in the latter
case~\cite{Priezjev18tp}.   Furthermore, with increasing strain
amplitude, the porous systems relocate to lower levels of the
potential energy, indicating structural changes in the glass phase
and pore redistribution (discussed below).  The effect of cyclic
loading on the lowest energy levels for different strain amplitudes
is summarized in the inset to Fig.\,\ref{fig:poten_cycle_number}.

\vskip 0.05in

\begin{figure}[t]
\includegraphics[width=12.cm,angle=0]{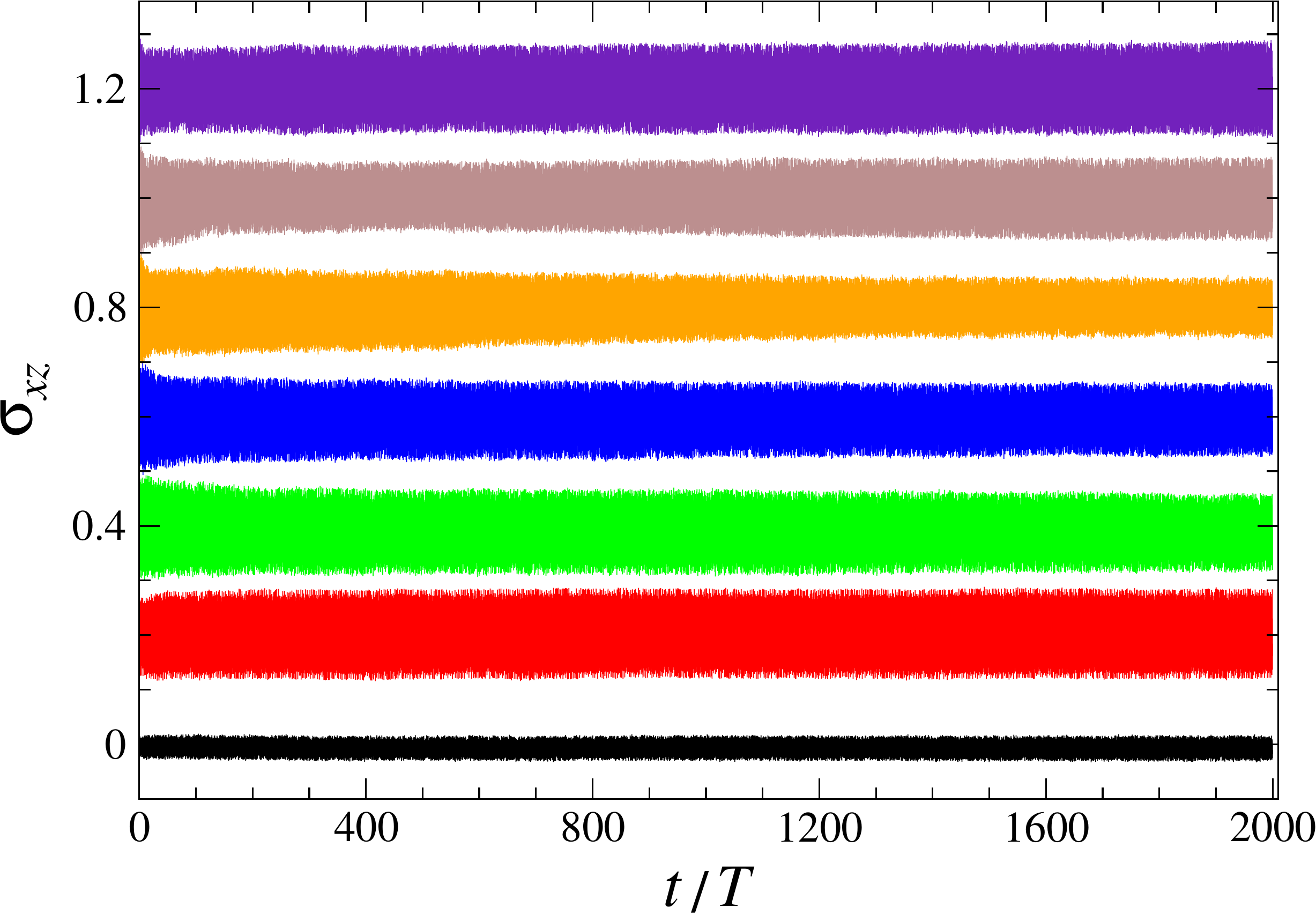}
\caption{(Color online) The time dependence of the shear stress
$\sigma_{xz}$ (in units of $\varepsilon\sigma^{-3}$) for strain
amplitudes  $\gamma_0=0.01$, $0.04$, $0.06$, $0.07$, $0.08$, $0.10$,
and $0.12$ (from bottom to top). For clarity, the data are displaced
upward by $0.2\,\varepsilon\sigma^{-3}$ for $\gamma_0=0.04$, by
$0.4\,\varepsilon\sigma^{-3}$ for $\gamma_0=0.06$, by
$0.6\,\varepsilon\sigma^{-3}$ for $\gamma_0=0.07$, by
$0.8\,\varepsilon\sigma^{-3}$ for $\gamma_0=0.08$, by
$1.0\,\varepsilon\sigma^{-3}$ for $\gamma_0=0.10$, and by
$1.2\,\varepsilon\sigma^{-3}$ for $\gamma_0=0.12$.  }
\label{fig:stress_cycle_2000}
\end{figure}
\begin{figure}[t]
\includegraphics[width=12.cm,angle=0]{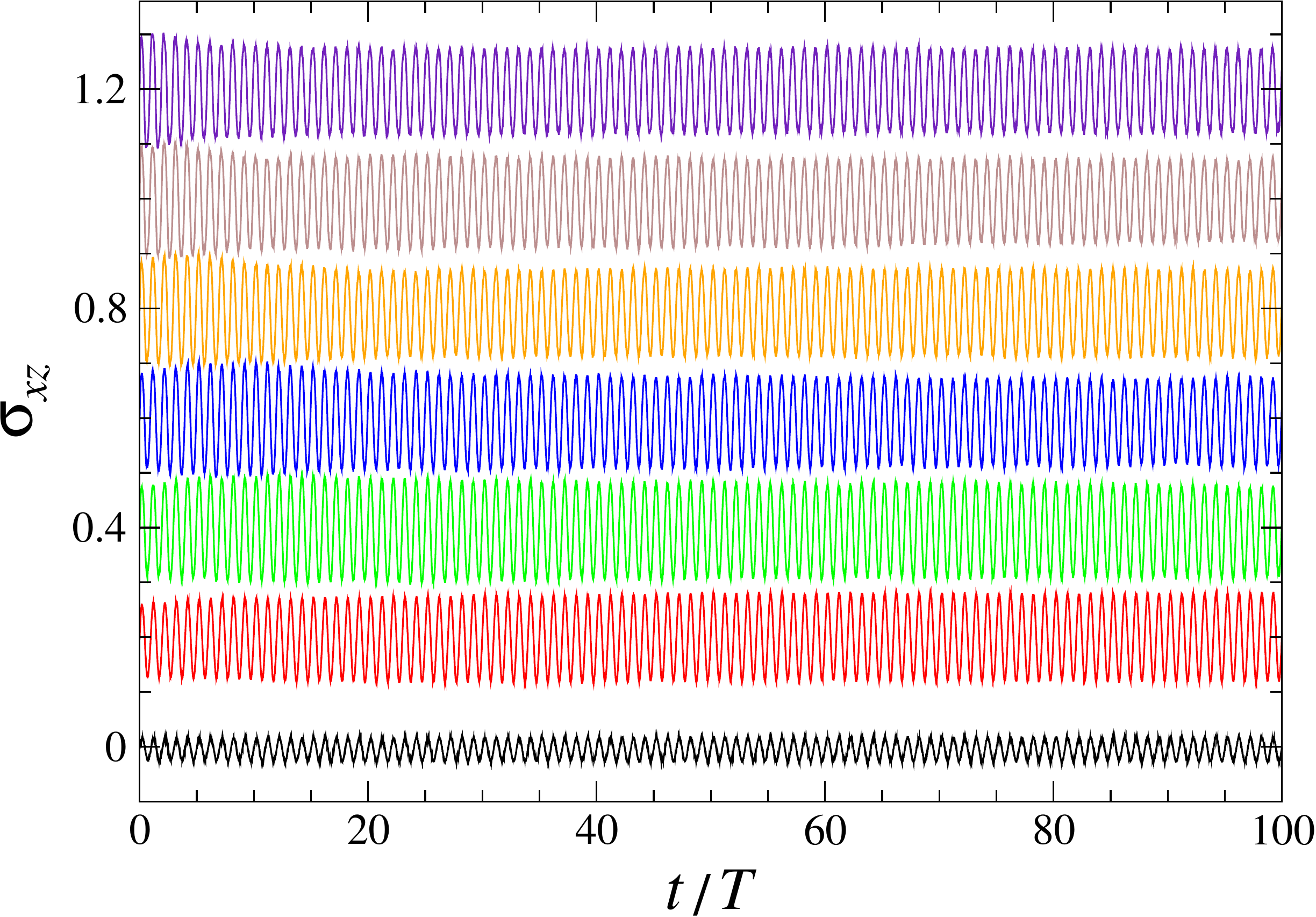}
\caption{(Color online) The shear stress $\sigma_{xz}$ (in units of
$\varepsilon\sigma^{-3}$) as a function of time for strain
amplitudes $\gamma_0=0.01$, $0.04$, $0.06$, $0.07$, $0.08$, $0.10$,
and $0.12$ (from bottom to top). The same data as in
Fig.\,\ref{fig:stress_cycle_2000} but plotted for the first 100
cycles. }
\label{fig:stress_cycle_100}
\end{figure}

Next, the shear stress during 2000 cycles is shown in
Fig.\,\ref{fig:stress_cycle_2000} for selected values of the strain
amplitude. Note that the data are shifted vertically for clarity. It
can be observed that after about a hundred transient cycles, the
amplitude of stress oscillations only weakly depends on the cycle
number.  Somewhat surprisingly, the stress amplitude after about 400
cycles slightly increases with time for $\gamma_{0}=0.10$, while it
steadily decreases for $\gamma_{0}=0.08$.  It can also be seen that
with increasing $\gamma_{0}$, the amplitude of stress oscillations
increases up to $\gamma_{0}=0.04$, indicating yielding and plastic
flow during oscillations with the strain amplitude $\gamma_{0}=0.06$
and above.  We comment that determination of the exact yielding
point and identification of the strain localization is not the main
focus of the study, as we examine long-term structural changes in
the glass phase and porous network.   Finally, a more detailed view
of the stress-strain response during the first 100 cycles is
presented in Fig.\,\ref{fig:stress_cycle_100}.  The local maximum in
stress amplitude during the first few cycles for
$\gamma_{0}\geqslant0.07$ is reminiscent to the behavior observed
for poorly annealed, homogeneous glasses cycled with the strain
amplitude $\gamma_{0}=0.06$, when the system was deformed nearly
reversibly with a relatively large stress amplitude for about a
hundred cycles before yielding and formation of a shear
band~\cite{Priezjev18a}.  As expected, the typical amplitude of
stress oscillations for the porous glass at a given $\gamma_{0}$ is
significantly lower than the stress amplitude for homogeneous binary
glasses with the average density $\rho\sigma^{3}=1.2$ reported in
the previous studies~\cite{Priezjev17,Priezjev18,Priezjev18a}.


\vskip 0.05in

\begin{figure}[t]
\includegraphics[width=15.cm,angle=0]{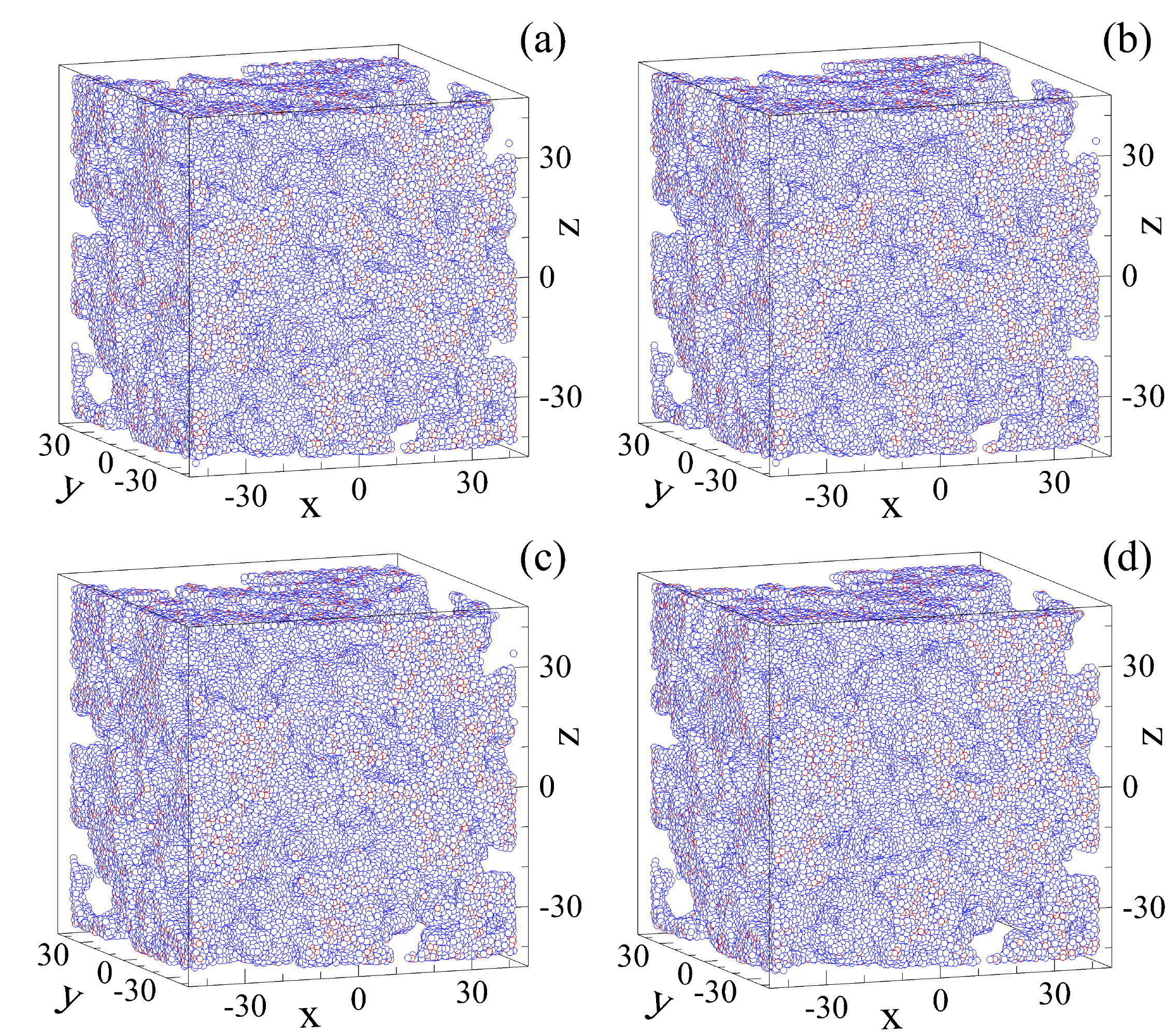}
\caption{(Color online) Instantaneous snapshots of the quiescent
porous glass ($\gamma_0=0$) after time intervals (a) $T$, (b)
$10\,T$, (c) $100\,T$, and (d) $2000\,T$. The period of oscillation
is $T=500\,\tau$ and the average glass density is
$\rho\sigma^{3}=0.5$. }
\label{fig:snapshot_gamma0_00}
\end{figure}
\begin{figure}[t]
\includegraphics[width=15.cm,angle=0]{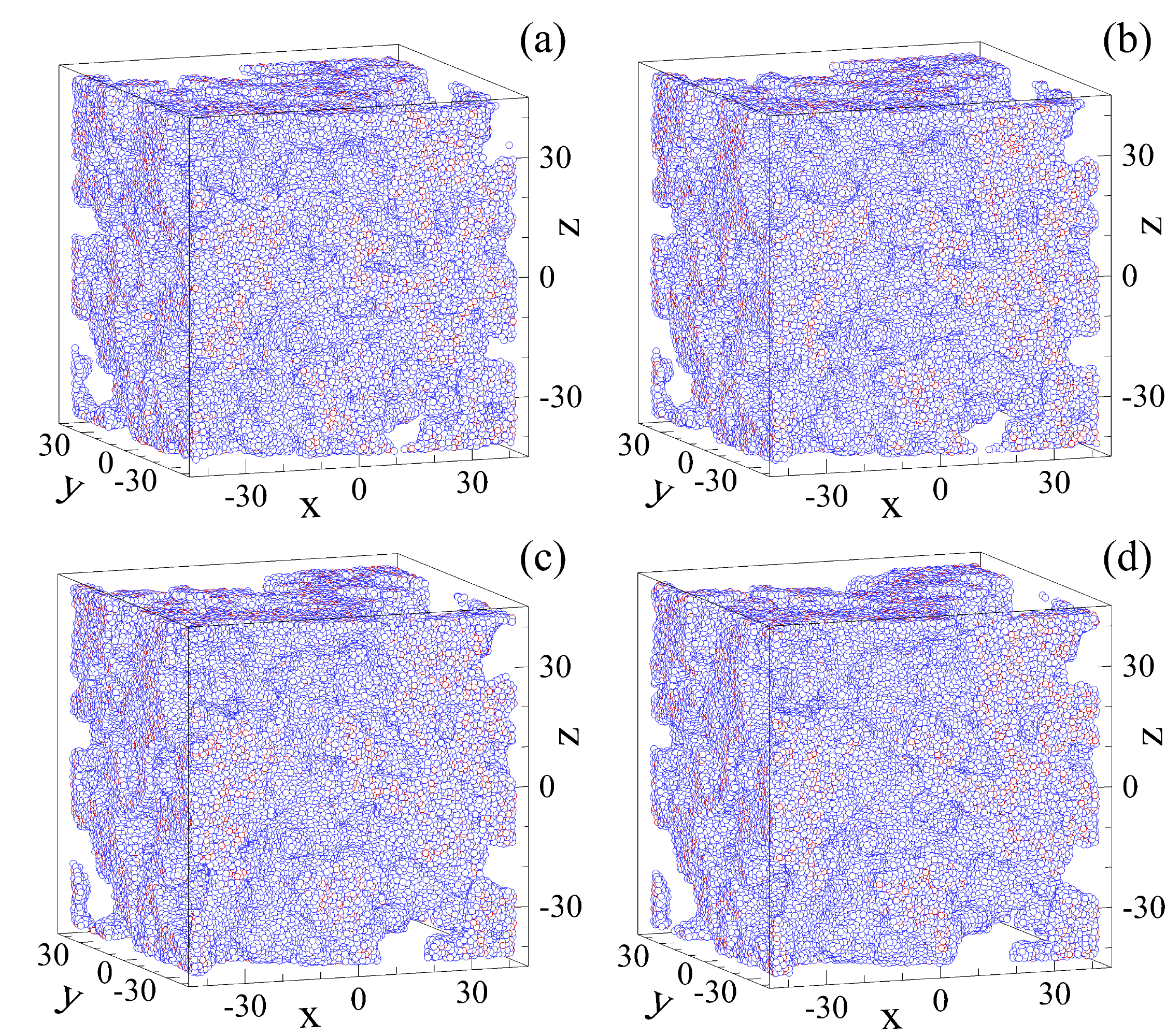}
\caption{(Color online) Atomic configurations of the porous glass
under oscillatory shear with the strain amplitude $\gamma_0=0.04$
after (a) 1-st, (b) 10-th, (c) 100-th, and (d) 2000-th cycle.}
\label{fig:snapshot_gamma0_04}
\end{figure}
\begin{figure}[t]
\includegraphics[width=15.cm,angle=0]{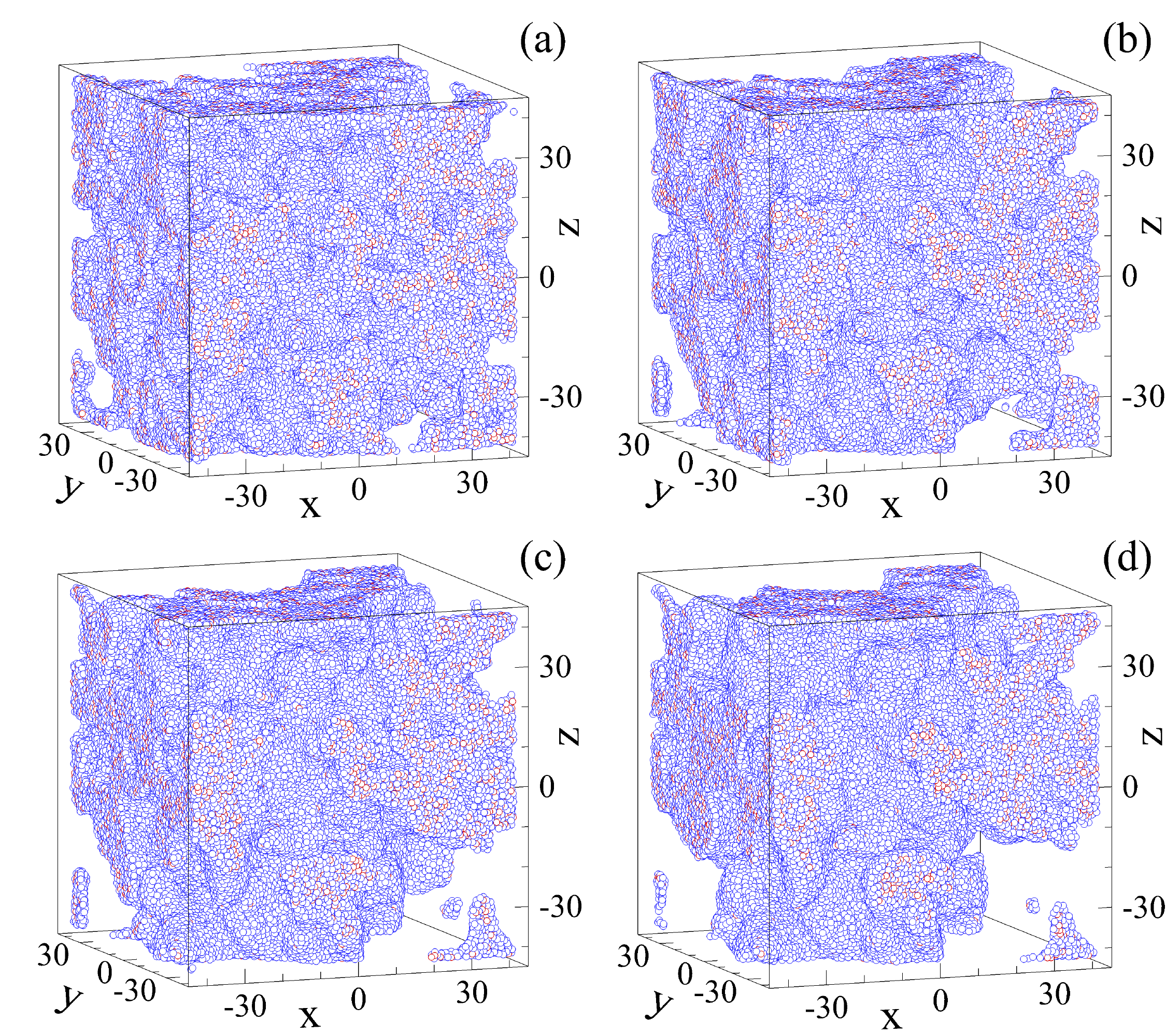}
\caption{(Color online) Instantaneous snapshots of the porous glass
periodically deformed with the strain amplitude $\gamma_0=0.08$
after (a) 1-st, (b) 10-th, (c) 100-th, and (d) 2000-th cycle.}
\label{fig:snapshot_gamma0_08}
\end{figure}
\begin{figure}[t]
\includegraphics[width=15.cm,angle=0]{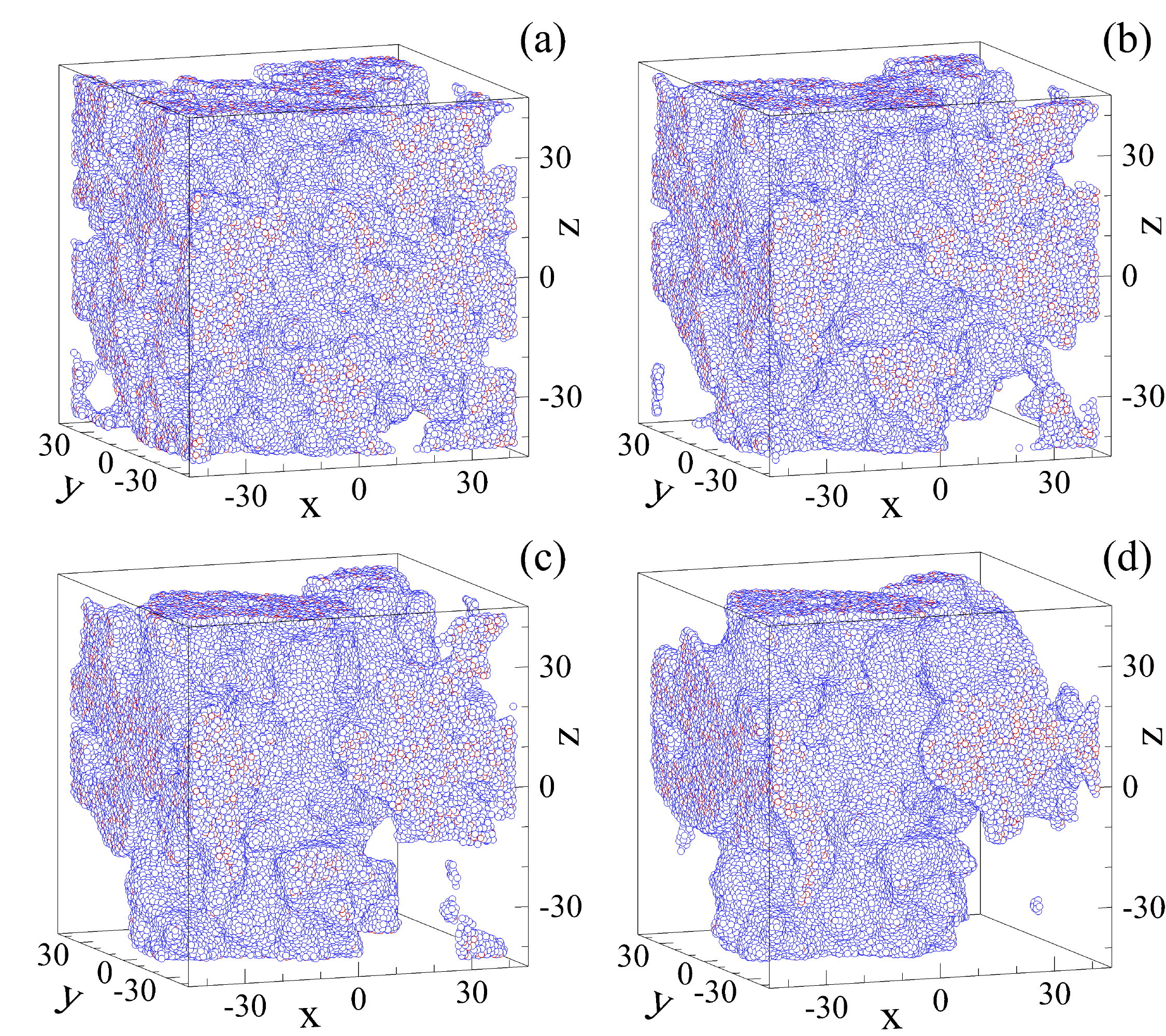}
\caption{(Color online) Atom positions during cyclic loading with
the strain amplitude $\gamma_0=0.12$ after (a) 1-st, (b) 10-th, (c)
100-th, and (d) 2000-th cycle.}
\label{fig:snapshot_gamma0_12}
\end{figure}

The representative snapshots of the porous glasses cycled with
different strain amplitudes are displayed in
Figs.\,\ref{fig:snapshot_gamma0_00}-\ref{fig:snapshot_gamma0_12}.
The atomic configurations are presented after the first, tenth,
100-th and the last cycle at zero strain.   It can be seen that the
porous structure in the quiescent glass, shown in
Fig.\,\ref{fig:snapshot_gamma0_00}, is composed of small isolated
pores and extended, highly tortuous channels.  During the aging
process at constant volume, the shape of the pores is only slightly
changed even after the longest time interval $t=2000\,T$, as shown
in Fig.\,\ref{fig:snapshot_gamma0_00}.  It can be further observed
that with increasing strain amplitude, the evolution of the porous
network becomes more apparent as larger pores are formed. Moreover,
at the largest strain amplitude $\gamma_0=0.12$, the small-size
pores become essentially expelled from the glass phase and solid
strands are formed across the system due to periodic boundary
conditions and the constraint of constant volume (see
Fig.\,\ref{fig:snapshot_gamma0_12}).  These structural changes in
periodically driven porous glasses can be quantified by computing
the distribution of spheres with different diameters that fill the
porous network.

\vskip 0.05in


The evaluation of the pore size distribution functions was performed
using the open-source Zeo++ software developed at the Lawrence
Berkeley National
Laboratory~\cite{Haranczyk12c,Haranczyk12,Haranczyk17}. More
specifically, the total volume of the system was decomposed into
Voronoi cells utilizing the VORO++ software
library~\cite{Rycroft09}.   As a results, the space around each atom
is surrounded by a polyhedral cell with edges that are equidistant
from neighboring atoms in a periodic simulation domain. Among other
structural characteristics, the probe accessible regions of the
void-space network can be identified using a modification of the
Dijkstra shortest path algorithm~\cite{Dijkstra59}.   In the present
study, the number of samples per atom is 50000 and the probe radius
is set to $0.3\,\sigma$.

\vskip 0.05in

\begin{figure}[t]
\includegraphics[width=12.cm,angle=0]{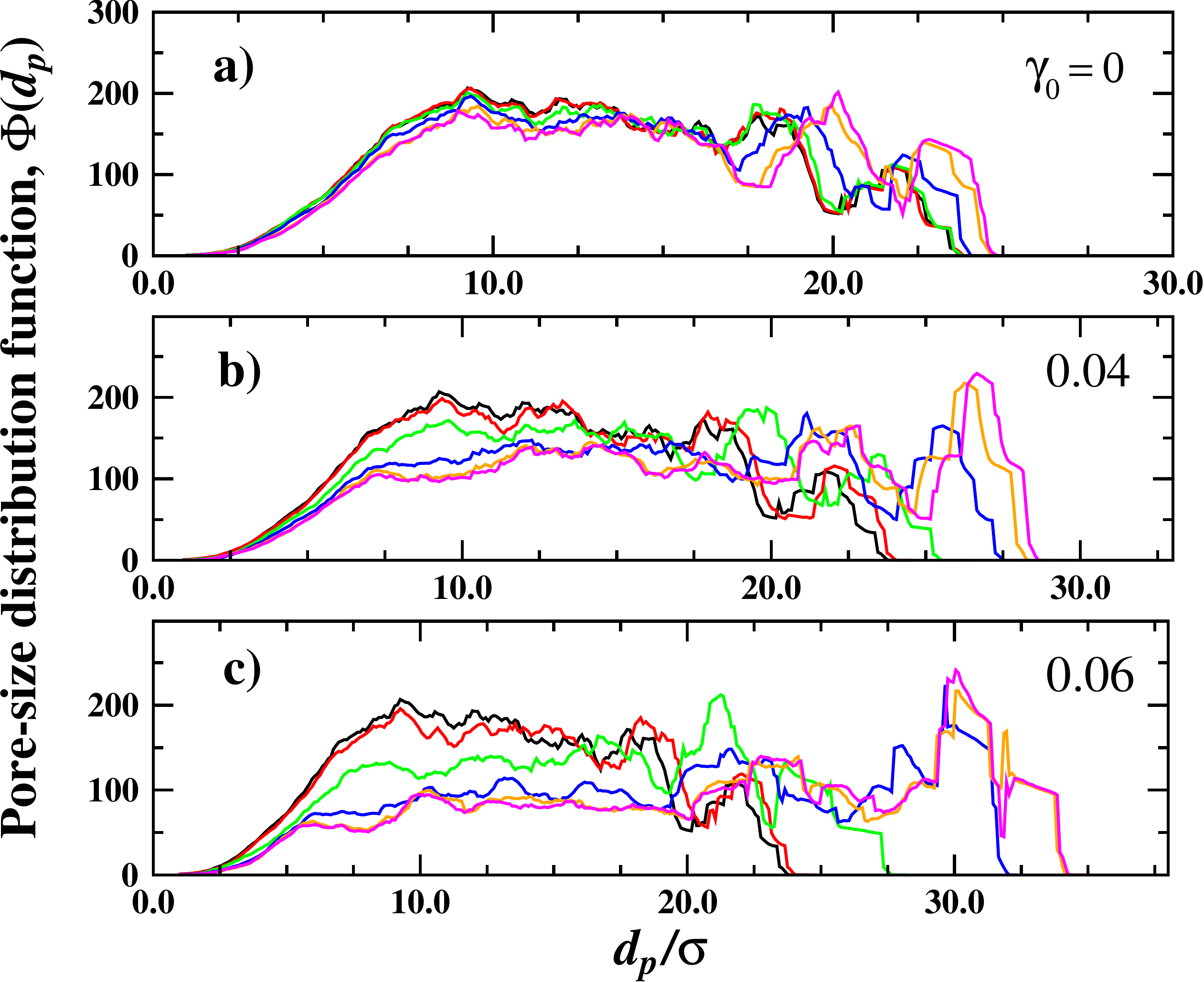}
\caption{(Color online) The pore size distribution (PSD) functions
for the strain amplitudes (a) $\gamma_0=0$, (b) $\gamma_0=0.04$, and
(c) $\gamma_0=0.06$. The colorcode for the cycle number is as
follows: 0 (black), 1 (red), 10 (green), 100 (blue), 1000 (orange),
and 2000 (magenta).}
\label{fig:pore_size_strain_amp_06}
\end{figure}
\begin{figure}[t]
\includegraphics[width=12.cm,angle=0]{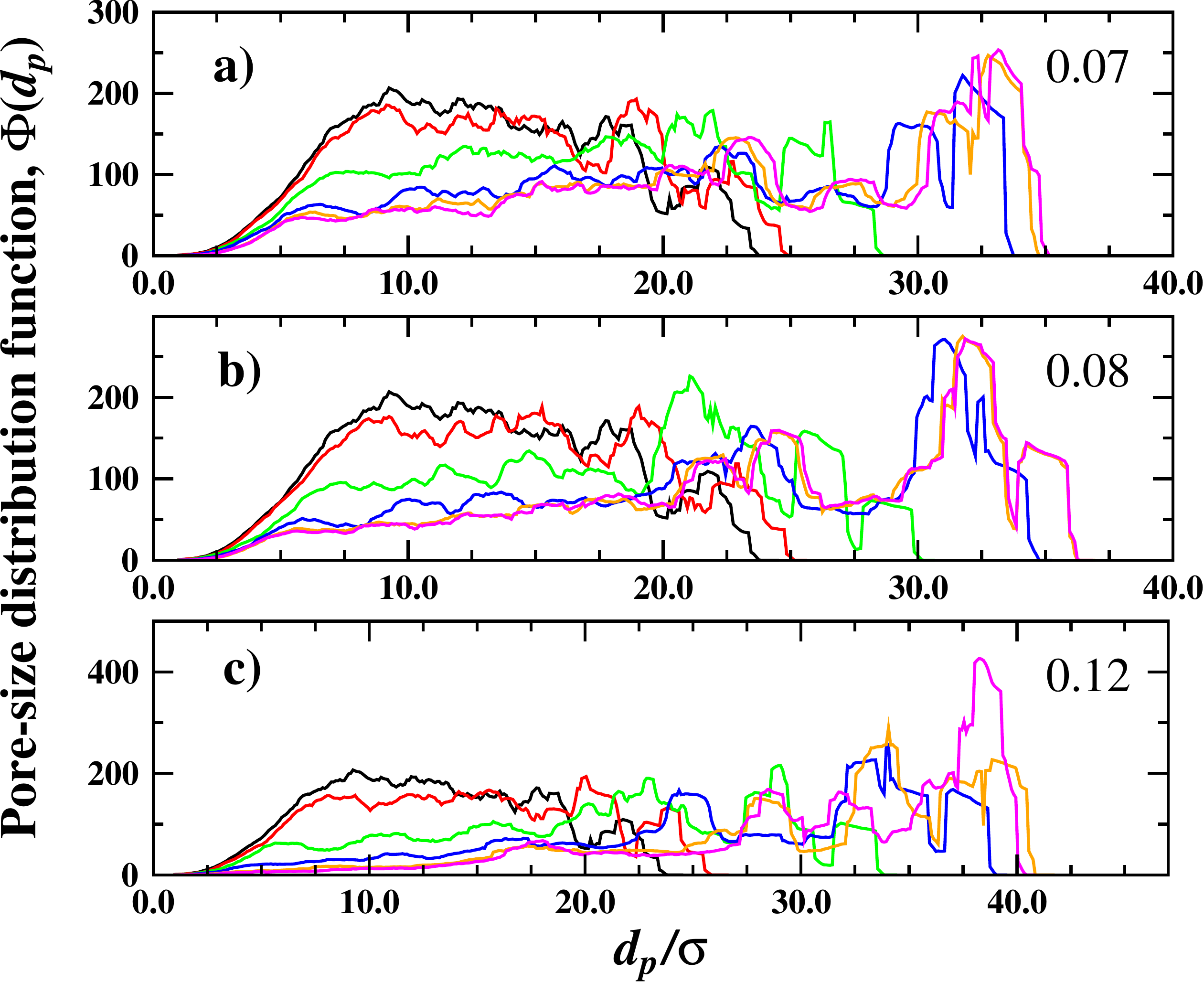}
\caption{(Color online) The pore size distribution functions for the
strain amplitudes (a) $\gamma_0=0.07$, (b) $\gamma_0=0.08$, and (c)
$\gamma_0=0.12$. The colorcode for the cycle number is 0 (black), 1
(red), 10 (green), 100 (blue), 1000 (orange), and 2000 (magenta).}
\label{fig:pore_size_strain_amp_12}
\end{figure}

The pore size distribution functions are presented in
Fig.\,\ref{fig:pore_size_strain_amp_06} for strain amplitudes
$\gamma_0=0$, 0.04, 0.06 and in
Fig.\,\ref{fig:pore_size_strain_amp_12} for $\gamma_0=0.07$, 0.08,
and 0.12.  In each case, the data were collected after the indicated
number of cycles at zero strain.   Note that the black curve denotes
the pore size distributions in the porous glass right after the
thermal quench ($t=0$) and it is the same in all panels in
Figs.\,\ref{fig:pore_size_strain_amp_06} and
\ref{fig:pore_size_strain_amp_12}.   It can be seen from
Fig.\,\ref{fig:pore_size_strain_amp_06}\,(a) that in the quiescent
glass, the distribution functions remain nearly the same during the
time interval $t=10\,T$, while larger pores are gradually developed
after the waiting time $t=2000\,T$. The effect is rather subtle and
it can hardly be detected by visual examination of the atomic
configurations shown in Fig.\,\ref{fig:snapshot_gamma0_00}.  By
contrast, in deformed glasses, the change in PSD functions becomes
apparent even after the first shear cycle, and the difference
between curves at $t=0$ and $t=T$ is amplified with increasing
strain amplitude.   Moreover, a pronounced peak at increasingly
large length scales is developed at larger strain amplitudes, and,
in addition, the peak height becomes larger with increasing either
cycle number or strain amplitude.  The largest transformation of the
porous structure can be observed for the strain amplitude
$\gamma_0=0.12$ shown in Fig.\,\ref{fig:snapshot_gamma0_12}, where a
large void space is formed around a compact glass phase after 2000
cycles. The appearance of such system-spanning void is reflected in
the highest peak at $d_p\approx 38\,\sigma$ shown in
Fig.\,\ref{fig:pore_size_strain_amp_12}\,(c).

\vskip 0.05in


To reveal a more complete picture of a porous material response to
periodic loading, in what follows, we focus on the analysis of the
local glass density.  Specifically, we compute the local density
distribution functions in the solid-state domains. The local
density,  $\langle\rho\rangle_R$, is defined as the number of atoms
located within the given radial range centered on a site of the
cubic lattice $L\in R^3$.  An analytical expression for the quantity
can readily be built using the following procedure. First, for each
lattice site, we define a closed ball, $B_R=\{ R \in \mathbb{R}^3,
\sum_{j=1}^3 R_j^2\leqslant R_0^2 \}$, where $R_0=|\vec{R}_0|$ is a
fixed non-zero rational number. Then, the on-site local density for
an atomic ensemble consisting of $N$ atoms can readily be obtained
as $\langle\rho\rangle_R=1/B_R \int dR^3 \delta(\vec{r}_i-\vec{R})$,
where the integral is taken over $B_R$ and $i=1, 2, ..., N$ is the
atomic index.   Note that  $\langle\rho\rangle_R$  can be regarded
as a measure of deviation of the local density from the average
density of a homogeneous dense glass.    Following the previous
analysis of porous glasses in a wide range of average glass
densities~\cite{Makeev18}, we used a fixed value $R_0=2.5\,\sigma$
in the present study.

\vskip 0.05in

\begin{figure}[t]
\includegraphics[width=12.cm,angle=0]{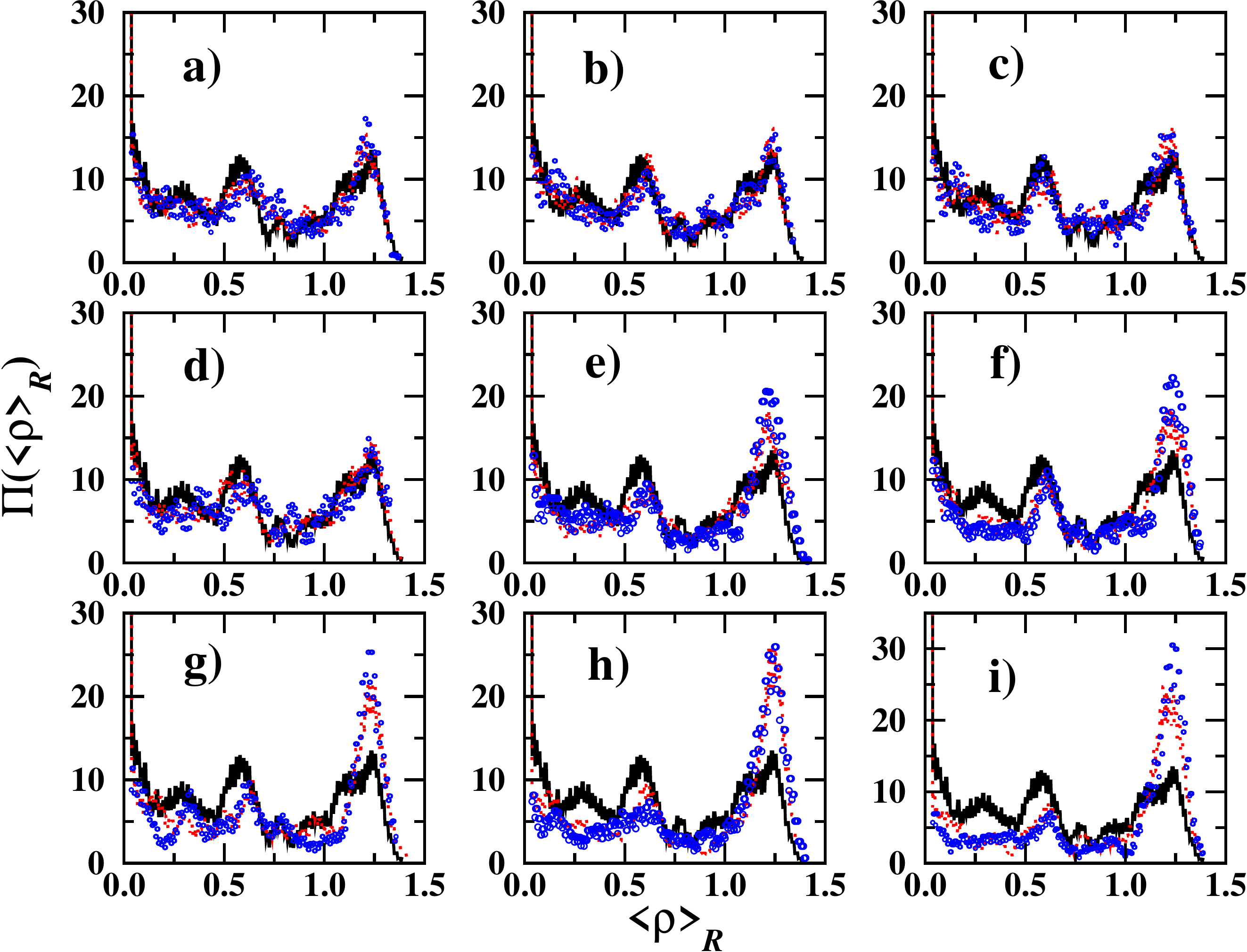}
\caption{(Color online) On-lattice local density distribution
functions, $\Pi(\langle\rho\rangle_R)$, for the selected strain
amplitudes, $\gamma_0$: (a) 0.0, (b) 0.01, (c) 0.02, (d) 0.04, (e)
0.06, (f) 0.07, (g) 0.08, (h) 0.10, and (i) $0.12$.   The
distributions were computed using the bin size
$\langle\rho\rangle_R^{\text{max}}/400$ and each set of data was
averaged over 20 data points for clarity. In each panel, we show the
density distribution functions computed at loading cycle number: $0$
(solid black line), $100\,T$ (dashed red line), and $2000\,T$ (open
blue circles). }
\label{fig:density_dist}
\end{figure}

In Fig.\,\ref{fig:density_dist}, we plot the local density
distribution functions, computed for porous glasses loaded with
strain amplitudes in the range $0  \leqslant \gamma_0 \leqslant
0.12$.   As can be deduced from Fig.\,\ref{fig:density_dist}, the
periodic loading causes substantial changes in the structural
properties of the solid domains.   The effects differ for smaller
density domains ($\rho\sigma^{3}<0.5$) and denser domains, in
particular, those where $\rho\sigma^{3}$ is close to the
corresponding density of a homogeneous glass. These structural
changes also depend strongly on the strain amplitude and cycle
number.    In particular, a significant homogenization of the solid
domains occurs in the solid regions with smaller density. On
average, these regions become less abundant in the driven systems.
Correspondingly, the intensity of the peak in the distribution
functions, located near the average density of dense non-porous
glassy phase, increases (see regions $\rho\sigma^{3}>1.0$ in
Fig.\,\ref{fig:density_dist}).   The effect is most pronounced for
the largest strain amplitude, $\gamma_0=0.12$, in which case the
increase in the peak value can be as large as in excess of 150\%. By
contrast, in the lower density regions, a flattening of the density
distribution functions and a decrease in their magnitudes is
observed. As clear from Fig.\,\ref{fig:density_dist}, the degree of
flattening increases with strain amplitude in the whole region of
strain amplitude variations.   Overall, we find that periodic
loading leads to significant homogenization of porous glasses,
wherein a growth of solid domains with densities characteristic for
dense non-porous glasses is observed. Further studies should focus
on the effects of periodic loading on the smaller-scale defects in
glasses. A work in this direction is currently in progress.

\section{Conclusions}

In summary, we performed a molecular dynamics simulation study of
porous glasses subjected to cyclic mechanical loading with varied
strain amplitude. The main emphasis of the study was put on temporal
evolution of the void spaces and solid domains in response to
periodic shear with varied amplitude and fixed period. We found that
cyclic loading causes substantial rearrangements in the structures
of both void spaces and solid domains. In particular, we observed
that there exist a segregation of void space and solid domains, such
that large pores are energetically favored. The effect is an
increasing function of the strain amplitude. Moreover, depending on
the strain amplitude, the periodic shear is found to considerably
facilitate both homogenization of the solid domains in porous
systems and leads to solid-domain densification, as revealed by the
behavior of the local density distributions, reported on in this
work. The combination of these observations allows the authors to
conclude that periodic mechanical loading can potentially be
employed as means for improving the structural properties of
metallic glasses and/or for design of glassy materials with desired
properties.

\section*{Acknowledgments}

Financial support from the National Science Foundation (CNS-1531923)
is gratefully acknowledged.  The article was prepared within the
framework of the Basic Research Program at the National Research
University Higher School of Economics (HSE) and supported within the
framework of a subsidy by the Russian Academic Excellence Project
`5-100'.  The molecular dynamics simulations were performed using
the LAMMPS numerical code~\cite{Lammps}. The distributions of pore
sizes were computed using the open-source software
ZEO++~\cite{Haranczyk12c,Haranczyk12,Haranczyk17}. Computational
work in support of this research was performed at Wright State
University's High Performance Computing Facility and the Ohio
Supercomputer Center.


%

%

%

%

%

%

%

%

%

%

\bibliographystyle{prsty}

\end{document}